\documentclass[aps,preprint]{revtex4}%
\usepackage{amsfonts}
\usepackage{amsmath}
\usepackage{amssymb}
\usepackage{graphicx}%
\providecommand{\U}[1]{\protect\rule{.1in}{.1in}}

\begin{document}
\preprint{ }
\title[Short title for running header]{Non-compact Groups, Coherent States, Relativistic\\Wave Equations and the Harmonic Oscillator II: Physical and geometrical considerations}
\author{Diego Julio Cirilo-Lombardo}
\affiliation{Bogoliubov Laboratory of Theoretical Physics}
\affiliation{Joint Institute for Nuclear Research, 141980, Dubna, Russian Federation.}
\email{diego@thsun1.jinr.ru ; diego77jcl@yahoo.com}
\author{}

\begin{abstract}
The physical meaning of the particularly simple non-degenerate supermetric,
introduced in the previous part by the authors, is elucidated and the possible
connection with processes of topological origin in high energy physics is
analyzed and discussed. New possible mechanism of the localization of the
fields in a particular sector of the supermanifold is proposed and the
similarity and differences with a 5-dimensional warped model are shown. The
relation with gauge theories of supergravity based in the $OSP(1/4)$ group is
explicitly given and the possible original action is presented. We also show
that in this non-degenerate super-model the physic states, in contrast with
the basic states, are observables and can be interpreted as tomographic
projections or generalized representations of operators belonging to the
metaplectic group $Mp\left(  2\right)  .$The advantage of geometrical
formulations based on non-degenerate super-manifolds over degenerate ones is
pointed out and the description and the analysis of some interesting aspects
of the simplest Riemannian superspaces are presented from the point of view of
the possible vacuum solutions .

\end{abstract}
\volumeyear{year}
\volumenumber{number}
\issuenumber{number}
\eid{identifier}
\date[Date text]{date}
\received[Received text]{date}

\revised[Revised text]{date}

\accepted[Accepted text]{date}

\published[Published text]{date}

\startpage{0}
\endpage{ }
\maketitle
\tableofcontents

\section{Introduction and motivation}

The study of the symmetries plays a fundamental role in modern physics. The
geometrical interpretation of the physical phenomena takes as basic object the
action, where all the dynamics of the theory is derived. The idea of to
associate an underlying geometrical structure to these physical phenomena
coming from of a fundamental idea of unification of all the interactions into
the natural world and not from an heuristic thought. The interrelation between
physical and mathematical definitions and concepts (i.e. geometry, groups,
topology$\leftrightarrow$space-time, internal structure, fields) turns more
and more concrete and basic in the physics of the XX and XXI centuries. If
well there are elegant formulations of the physical problems of interest from
the mathematical point of view, there exists a lack of uniqueness in the
geometrical definition of the Lagrangian density.

The great difficulties appear (or almost are evidently explicit) at the
quantum level where the geometrical objects playing the role of Lagrangian or
Hamiltonian pass to play the role of (super) operators. These troubles carry
inexorably to the utilization of diverse methods or prescriptions that change
the original form of the action (or Hamiltonian). This distortion of the
original form of these fundamental operators at the classical level generally
does not produces changes into the dynamical equations of the theory but
quantically introduces several changes, because the spectrum of physical
states is closely related with the form of the Hamiltonian. This fact was
pointed out by the authors in the previous paper [10]. Clearly, in order to
construct the Lagrangian and other fundamental invariants of the theory, the
introduction of a manifold as the important ingredient is the relevant thing.
In particular can be very interesting the introduction of a super-manifold (in
the sense of [10] and references therein) in order to include the fermionic
fields in a natural manner.

Several attempts have been made by various groups to construct the theory of
supergravity as the geometry of a superspace possessing non-zero curvature and
torsion tensors without undesirable higher spin states[1,2]. Only few years
after those works, the consistent construction of the superfield supergravity
was formulated in the pioneering papers independently by V.I. Ogievetsky and
E. Sokatchev [3] and S.J. Gates and W. Siegel [4]. From these times in several
areas of the theoretical physics the description of different systems was
given in the context of the geometry of supermanifolds and superfields[5]:
supergravity and d-branes models with warped supersymmetry[6], super-Landau
systems[7], superbrane actions from nonlinearly realized supersymmetries[8], etc.

It is therefore of interest to study the geometry not only of the simplest
superspaces, but also the more unusual or non-standard ones and elucidate all
the gauge degrees of freedom that they possesses. This fact will clarify and
expand the possibilities to construct more realistic physical models and new
mathematically consistent theories of supergravity. On the other hand, the
appearance of supergroups must draw attention to the study of the geometries
of the homogeneous superspaces whose groups of motions they are. Another
motivation of the study of these Riemannian superspaces is in order to
establish some degree of uniqueness in the obtained supersymmetric solutions.

Motivated by the above, we complete our previous paper [10] studying and
analyzing from the point of view of the possible vacuum solutions, the
simplest non trivial supermetric given by Volkov and Pashnev in [9] that was
the \textquotedblleft starting point\textquotedblright\ toy model of the first
part of this work%
\begin{equation}
ds^{2}=\omega^{\mu}\omega_{\mu}+\mathbf{a}\omega^{\alpha}\omega_{\alpha
}-\mathbf{a}^{\ast}\omega^{\overset{.}{\alpha}}\omega_{\overset{.}{\alpha}}
\tag{1}%
\end{equation}
This particular non-degenerate supermetric contains the complex parameters
$\mathbf{a}$ and $\mathbf{a}^{\ast}$ that make it different of other more
standard supermetrics. As we shown in [10,11], the degenerate supermetrics are
not consistent into a well theoretically formulated supergeometry. Then, our
main task is to find the meaning and the role played by these complex
parameters from the geometrical and physical point of view. To this end, we
compare the solution of ref.[10] that was computed in the N=1 four dimensional
superspace proposed in [9,11], compactified to one dimension and restricted to
the pure time-dependent case with:

i) the well known solution described in references[12,13] that was formulated
in a superspace $\left(  1\mid2\right)  .$

ii) a multidimensional warped model described in [14], in this case also
considering for the proposed superspace the possible dependence of the
solution on $n-$additional bosonic coordinates (d=n+4).

Our goal is to show that, from the point of view of the obtained solutions,
the complex parameters $\mathbf{a}$ \textit{localize} the fields in a specific
region of the bosonic part of this special superspace, they explicitly
\textit{breakdown }the chiral symmetry when some conditions are required and
all these very important properties \textit{remain} although the supersymmetry
of the model was completely broken. Also, besides all these highlights, we
also show that the obtained vacuum states from the extended supermetric are
very well defined in any Hilbert space.

About the geometrical origin of this particularly special metric, we
demonstrate that it can be \textit{naturally derived} from a theory of
supergravity based in a $OSP(1/4)$-valuated connection $A$. \ When the
symmetry of the $OSP(1/4)$ model (super-$SO(4,1)$) presented here is
explicitly written as a function of its reductive components, a part as (1)
appear plus a first order (Dirac-like) fermionic term.

And finally we show that the \textit{physical states} derived of the
geometrical Lagrangian constructed with this particular supermetric are
nothing more that the \textit{tomographic projections} (in the sense given by
the authors of [34]) of the Heisenberg-Weyl and $su(1,1)$ fundamental
operators (previously defined in [10]) $L_{ab}=$ $\left(
\begin{array}
[c]{c}%
a\\
a^{+}%
\end{array}
\right)  _{ab}$ \ $\ $and $\mathbb{L}_{ab}=$ $\left(
\begin{array}
[c]{c}%
a^{2}\\
a^{+2}%
\end{array}
\right)  _{ab}$ with respect to the basic coherent states (CS)
\textit{fundamental solutions} of the square root of the interval (1). These
physical states have the following spin content: $\lambda=1/2,1,3/2$ and 2 and
the representations of these fundamental operators are in \textit{diagonal
(Sudarshan-like)} and in an \textit{asymmetric-(non diagonal)} form, both
representations forming part of a more general class of representations for
operators recently proposed by Klauder and Skagerstam in ref.[35].

The plan of the paper is as follows: in Section 2 we give a brief review,
based in a previous work of the author [10], about the $N=1$ non-degenerate
four dimensional superspace proposed by Volkov and Pashnev and its solution.
Section 3 is devoted to analyze the relation of the supermetric under
consideration with the superspace $\left(  1\mid2\right)  ~\ $given explicitly
under which conditions one is reduced to the other one from the point of view
of the obtained vacuum solutions. The geometrical derivation of the
supermetric from a gauge theory of supergravity based in the $OSP(1/4)$ group
(super-$SO(4,1)$), a \textit{new} superparticle model and the link between the
complex parameters $\mathbf{a}$ and $\mathbf{a}^{\ast}$ and the cosmological
constant $\Lambda$ are given in Section 4. In Section 5 a surprising
connection between the extended supermetric and multidimensional warped
gravity model solutions is shown and some hints of a possible \textit{new
mechanism} of the field localization and the idea of confinement is proposed.
In Section 6 and 7 the interpretation of the physic states of the theory as
\textit{tomographic representations} of operators of the metaplectic group
Mp$\left(  2\right)  $ are discussed, the spin content of these states is
analyzed and the Gram-Schmidt operator is explicitly given. Finally in Section
8 we resume the main results and concluding remarks.

\section{The particular four dimensional $N=1$ superspace}

The superspace $\left(  1,3\mid1\right)  $ has four bosonic coordinates
$x^{\mu}$and one Majorana bispinor$:\left(  t,x^{i},\theta^{\alpha}%
,\overline{\theta}^{\overset{.}{\alpha}}\right)  $. Two possible realizations
for this superspace are$\rightarrow\left\{
\begin{array}
[c]{c}%
osp\left(  2,2\right)  \rightarrow Bosonic-Fermionic\\
osp\left(  1/2,\mathbb{R}\right)  \rightarrow Bosonic
\end{array}
\right.  $ with the following group structure for the bosonic-fermionic
realization%
\[
\left(
\begin{array}
[c]{cc}%
SU\left(  1,1\right)  & Q\\
Q & SU(1,1)
\end{array}
\right)
\]
We will concentrate our analysis to the superspace $\left(  1,3\mid1\right)  $
with extended line element as in [9,11]
\begin{equation}
ds^{2}=\omega^{\mu}\omega_{\mu}+\mathbf{a}\omega^{\alpha}\omega_{\alpha
}-\mathbf{a}^{\ast}\omega^{\overset{.}{\alpha}}\omega_{\overset{.}{\alpha}}
\tag{1}%
\end{equation}
invariant to the following supersymmetric transformations%

\[
x_{\mu}^{\prime}=x_{\mu}+i\left(  \theta^{\alpha}\left(  \sigma_{\mu}\right)
_{\alpha\overset{.}{\beta}}\overline{\xi}^{\overset{.}{\beta}}-\xi^{\alpha
}\left(  \sigma_{\mu}\right)  _{\alpha\overset{.}{\beta}}\overline{\theta
}^{\overset{.}{\beta}}\right)  \ ;\ \ \theta^{\prime\alpha}=\theta^{\alpha
}+\xi^{\alpha}\ ;\ \ \overline{\theta^{\prime}}^{\overset{.}{\alpha}%
}=\overline{\theta}^{\overset{.}{\alpha}}+\overline{\xi}^{\overset{.}{\alpha}}%
\]
where the Cartan forms of the group of the supersymmetry are%

\[
\omega_{\mu}=dx_{\mu}-i\left(  d\theta\ \sigma_{\mu}\overline{\theta}%
-\theta\ \sigma_{\mu}d\overline{\theta}\right)  ;\ \ \ \ \ \ \ \ \omega
^{\alpha}=d\theta^{\alpha}\ ;\ \ \ \ \ \ \ \ \omega^{\overset{.}{\alpha}%
}=\overline{d\theta}^{\overset{.}{\alpha}}%
\]
The spinorial indices are related as follows (the dotted indices are similarly
related, as usual):%
\[
\theta^{\alpha}=\varepsilon^{\alpha\beta}\theta_{\beta}%
\ \ \ \ \ ;\ \ \ \ \ \theta_{\alpha}=\theta^{\beta}\varepsilon_{\beta\alpha
}\ \ \ ;\ \ \varepsilon_{\alpha\beta}=-\varepsilon_{\beta\alpha}%
\ \ \ \ ;\ \ \ \varepsilon^{\alpha\beta}=-\varepsilon^{\beta\alpha
}\ \ \ ;\ \ \ \varepsilon_{12}=\varepsilon^{12}=1
\]
The complex constants $\mathbf{a}$ and $\mathbf{a}^{\ast}$ in the extended
line element are arbitrary. This arbitrarity for the choice of $\mathbf{a}$
and $\mathbf{a}^{\ast}$are constrained by the invariance and reality of the
interval (1). The solution for the metric in the time dependent case with 3
spatial dimensions compactified (i.e. $\mathbf{R}^{1}\otimes S^{3}$, ref.[15])
takes the form [10]
\begin{equation}
g_{ab}\left(  t\right)  =e^{A\left(  t\right)  +\xi\varrho\left(  t\right)
}g_{ab}\left(  0\right)  \tag{2}%
\end{equation}
with the following superfield solution
\[
\varrho\left(  t\right)  =\phi_{\alpha}+\overline{\chi}_{\overset{.}{\alpha}}%
\]
(i.e. chiral plus anti-chiral parts). The system of equations for $A\left(
t\right)  $ and $\varrho\left(  t\right)  $ that we are looking for was given
in [10], and is the following%
\begin{equation}%
\begin{array}
[c]{c}%
\left\vert \mathbf{a}\right\vert ^{2}\ddot{A}+m^{2}=0\\
\overset{..}{\overline{\chi}}_{\overset{.}{\alpha}}-i\frac{\omega}{2}\left(
\sigma^{0}\right)  _{\ \overset{.}{\alpha}}^{\alpha}\ \phi_{\alpha}=0\\
-\overset{..}{\phi}_{\alpha}+i\frac{\omega}{2}\left(  \sigma^{0}\right)
_{\alpha}^{\overset{.}{\ \beta}}\ \overline{\chi}_{\overset{.}{\beta}}=0
\end{array}
\tag{3}%
\end{equation}
The above system can be solved exactly given us the following result%
\begin{equation}
A=-\frac{1}{2}\left(  \frac{m}{\left\vert \mathbf{a}\right\vert }\right)
^{2}t^{2}+c_{1}t+c_{2}\ ;\ \ c_{1},c_{2}\in\mathbb{C} \tag{4}%
\end{equation}
and%
\begin{equation}
\phi_{\alpha}=\overset{\circ}{\phi}_{\alpha}\left(  \alpha e^{i\omega
t/2}+\beta e^{-i\omega t/2}\right)  +\frac{2i}{\omega}\left(  \sigma
^{0}\right)  _{\alpha}^{\overset{.}{\ \beta}}\ \overline{Z}_{\overset{.}%
{\beta}} \tag{5}%
\end{equation}%
\begin{equation}
\overline{\chi}_{\overset{.}{\alpha}}=\left(  \sigma^{0}\right)
_{\ \overset{.}{\alpha}}^{\alpha}\ \overset{\circ}{\phi}_{\alpha}\left(
\alpha e^{i\omega t/2}-\beta e^{-i\omega t/2}\right)  +\frac{2i}{\omega
}\left(  \sigma^{0}\right)  _{\ \overset{.}{\alpha}}^{\alpha}\ Z_{\alpha}
\tag{6}%
\end{equation}
where $\overset{\circ}{\phi}_{\alpha},\ Z_{\alpha}$ and $\overline
{Z}_{\overset{.}{\beta}}$ are constant spinors and the frequency goes as:
$\omega^{2}\sim\frac{4}{\left\vert a\right\vert ^{2}}$. The superfield
solution for the fields (see the "square states" of ref.[10,11]) that we are
looking for, have the following form%
\begin{equation}
g_{ab}\left(  t\right)  =e^{-\frac{1}{2}\left(  \frac{m}{\left\vert
\mathbf{a}\right\vert }\right)  ^{2}t^{2}+c_{1}t+c_{2}}e^{\xi\varrho\left(
t\right)  }g_{ab}\left(  0\right)  \tag{7}%
\end{equation}
with
\begin{equation}
\varrho\left(  t\right)  =\overset{\circ}{\phi}_{\alpha}\left[  \left(  \alpha
e^{i\omega t/2}+\beta e^{-i\omega t/2}\right)  -\left(  \sigma^{0}\right)
_{\ \overset{.}{\alpha}}^{\alpha}\ \left(  \alpha e^{i\omega t/2}-\beta
e^{-i\omega t/2}\right)  \right]  +\frac{2i}{\omega}\left[  \left(  \sigma
^{0}\right)  _{\alpha}^{\overset{.}{\ \beta}}\ \overline{Z}_{\overset{.}%
{\beta}}+\left(  \sigma^{0}\right)  _{\ \overset{.}{\alpha}}^{\alpha
}\ Z_{\alpha}\right]  \tag{8}%
\end{equation}
and%
\begin{equation}
g_{ab}\left(  0\right)  =\left\langle \Psi\left(  0\right)  \right\vert
L_{ab}\left\vert \Psi\left(  0\right)  \right\rangle \tag{9}%
\end{equation}
that is nothing more that the "square" of the state $\ \Psi^{\left[  1\right]
}$\footnotetext[1]{This particular realization was initially introduced in
ref.[28]) between the fundamental states $\left\vert \Psi\right\rangle $ in
the initial time, where the subalgebra is the Heisenberg-Weyl algebra (with
generators $a,$ $a^{+}$ and$\left(  n+\frac{1}{2}\right)  $)} ($L_{ab}=$
$\left(
\begin{array}
[c]{c}%
a\\
a^{+}%
\end{array}
\right)  _{ab}$ \ with $a$ and $a^{+}$ the standard creation and annihilation
operators). \ The meaning of the expression (9) was given by the authors in
ref.[10] and can be resumed as:

i) it can be interpreted as the "square" of the state $\ \Psi$ and it is the
\textit{fundamental solution} of the square root of the interval (1),
precisely describing a \textit{trajectory} in the superspace[9,10,11];

ii) for these states $\Psi$ the\textit{ zero component of the} \textit{current
is not positively definite} given explicitly by%
\[
j_{0}\left(  x\right)  =2E\Psi^{\dagger}\Psi
\]
but for the states $g_{ab}$%
\[
j_{0}\left(  x\right)  =2E^{2}g^{ab}g_{ab}%
\]
then, $j_{0}\left(  x\right)  $ for the states $g_{ab}$ is positively definite
(e.g. the energy $E$ appears squared);

iii) from ii), such states $\Psi$ are related with \textit{ordinary physical
observables} only through they "square" $g_{ab}$ in the sense of expressions
as (9), and this fact is very important in order to explain the reason why
these fractional spin states are not easily observed or detected in the nature
under ordinary conditions [10];

iv) and fundamentally we will take under consideration here only the
particular case of spin 2 because for this state the Hilbert space is dense
and these states lead a \textit{thermal} spectrum$^{[2]}$\footnotetext[2]{The
other possibilities are squeezed states (non-thermal spectrum).} [10,16](
$g_{ab}$ in the expression (9) has s=2: each state $\Psi$ contributes with a
spin weight equal to one, see detailed explanation in Section 6). Other
interesting possibilities given by these type of coherent states solutions and
they physical meaning, that can give some theoretical framework for more
degrees of freedom for the graviton in the sense of [29-31], will be analyzed
with details in a separate paper[16](see also Section 6).

The $g_{ab}$ at time t is given by the following expression[10, Appendix]%

\begin{equation}
g_{ab}\left(  t\right)  =e^{-\left(  \frac{m}{\left\vert \mathbf{a}\right\vert
}\right)  ^{2}t^{2}+c_{1}^{\prime}t+c_{2}^{\prime}}e^{\xi\varrho\left(
t\right)  }\left\vert f\left(  \xi\right)  \right\vert ^{2}\left(
\begin{array}
[c]{c}%
\alpha\\
\alpha^{\ast}%
\end{array}
\right)  _{ab} \tag{10}%
\end{equation}
where $\alpha$ and $\alpha^{\ast}$ are the respective eigenvalues of the
creation-annihilation operators $\ a$ and $a^{+}$. And the dynamics for $\Psi
$\ becomes now to%
\begin{equation}
\Psi_{\lambda}\left(  t\right)  =e^{-\frac{1}{2}\left[  \left(  \frac
{m}{\left\vert \mathbf{a}\right\vert }\right)  ^{2}t^{2}+c_{1}^{\prime}%
t+c_{2}^{\prime}\right]  }e^{\frac{\xi\varrho\left(  t\right)  }{2}}\left\vert
f\left(  \xi\right)  \right\vert \left(
\begin{array}
[c]{c}%
\alpha^{1/2}\\
\alpha^{\ast1/2}%
\end{array}
\right)  _{\lambda} \tag{11}%
\end{equation}
It's useful to remark here that there exist some misleadings and wrong
asseverations about the non-degenerate supermetrics as (1) in some references
(see e.g.:[38]). The reason of these wrong claims coming from the
misunderstanding that metrics as (1) in appearance don't give a Dirac's type
(1$^{st}$ order) equations of motion for fermions. In Section 5, when we
discuss the origin of this type of metrics, this fact will be completely clear.

\section{Relation with the $\left(  1\mid2\right)  $ superspace}

We pass now to the description of the superspaces under consideration from the
uniqueness of the possible solutions for the metric components, the supergroup
structures defined by the possible group of motions and the possible physical
interpretation of these results. The superspace $\left(  1,2\right)  $ has one
bosonic coordinate $t$ and two majorana spinors: $x^{\mu}\equiv\left(
t,\theta^{1},\theta^{2}\right)  $ and is the simplest low dimensional
superspace of interest (we use similar notation as in refs.[12,13]). The big
group in which this superspace is contained is $OSP\left(  3,2\right)  $,
schematically as%

\[
\left(
\begin{array}
[c]{cc}%
O\left(  3\right)  & Q\\
Q & SP\left(  2\right)
\end{array}
\right)
\]
The solution for the metric in this case is given by [12,13]%
\begin{equation}
\overline{g}_{ab}=g_{ab}e^{2\sigma\left(  t,\theta\right)  } \tag{12}%
\end{equation}
where the following superfield was introduced%
\[
\sigma\left(  t,\theta\right)  =A\left(  t\right)  +\theta^{\beta}B_{\beta
}+\theta^{\alpha}\theta_{\alpha}F\left(  t\right)
\]
From the Einstein equations for the $\left(  1\mid2\right)  $ superspace we
obtain the following set%
\begin{equation}
\left\{
\begin{array}
[c]{c}%
\overset{.}{B_{\alpha}}+b_{\alpha}^{\beta}B_{\beta}-\overset{.}{A}B_{\alpha
}=0\\
\overset{..}{A}-\frac{1}{2}\overset{.}{A^{2}}+\frac{1}{2}B^{\gamma}B_{\gamma
}=\frac{\lambda}{4}\left(  e^{2A}-1\right)
\end{array}
\right.  \tag{13}%
\end{equation}
where $b_{\alpha\beta}=$ $b_{\beta\alpha}$ is an arbitrary symmetric matrix.
Making a suitable transformation in the first of above equations the explicit
form of the $B_{\gamma}$ field that we are looking for is%
\begin{equation}
B^{\gamma}B_{\gamma}=\nu^{\alpha}\nu_{\alpha}e^{2A}\ \tag{14}%
\end{equation}
$\ \nu_{\alpha}$ is a constant spinor and $\sqrt{b}$ was associated in the
ref.[13] with the mass. Inserting (14) in the second equation of the system
(13) it leads the following new equation%
\begin{equation}
\overset{..}{A^{\prime}}-\frac{1}{2}\overset{.}{A^{\prime2}}=\frac{\lambda}%
{4}\left(  e^{2A^{\prime}}-1\right)  \tag{15}%
\end{equation}
where the transformation $A^{\prime}=A-\frac{\nu^{\alpha}\nu_{\alpha}}%
{\lambda}$ was used. Notice that in the ref.[13] the derivation of the
solution of the equation (15) was not explicitly explained, but however it is
easy to see that can be reduced to the following expression%
\begin{equation}
\left(  \overset{.}{W}\right)  ^{2}=\frac{\lambda}{4}\left(  W^{2}+\frac
{1}{2W^{2}}\right)  +C \tag{16}%
\end{equation}
with $W=e^{-\frac{A^{\prime}}{2}}$ and $C$ is an arbitrary constant. When
$C=0$ eq.(16) is the equation of motion for a supersymmetric oscillator in the
potential of the form $k\left(  X^{2}+\frac{1}{X^{2}}\right)  $, for which the
group $O\left(  3\right)  $ is a dynamic symmetry group$.$ Notice that from
the point of view of a potential it is possible redefine it in order that C
disappears, but the conservation of C is crucial for the determination of the
families of solutions of the problem (is not possible to know completely this
type of problems \textit{only} inspectioning the potential). This type of
equations of motion for an oscillator with conformal symmetry was considered
earlier in the non-supersymmetric case in [17]. The solutions for the possible
values of the constant $C$ are%
\[
C=0\rightarrow e^{-A}=\frac{\sqrt{2}}{2}Sinh\left(  \sqrt{\lambda}%
t+\varphi_{0}\right)  ,\ \ \ \ \ \ \ \ \ \ \ \varphi_{0}=\sqrt{\lambda}t_{0}%
\]%
\[
\frac{8C^{2}}{\lambda^{2}}<1\rightarrow e^{-A}=\frac{\sqrt{2}}{2}\left[
Sinh\left(  \sqrt{\lambda}t+\varphi_{0}\right)  \sqrt{1-\varkappa^{2}%
}-\varkappa\right]  ,\ \ \ \ \ \ \ \ \ \ \ \varkappa=\frac{2\sqrt{2}C}%
{\lambda}%
\]%
\[
\frac{8C^{2}}{\lambda^{2}}=1\rightarrow e^{-A}=\frac{\sqrt{2}}{2}\left[
\frac{e^{\left(  \sqrt{\lambda}t+\varphi_{0}\right)  }}{\sqrt{2}}%
-\varkappa\right]  ,\ \ \ \ \ \ \ \ \ \
\]%
\begin{equation}
\sigma\left(  t,\theta\right)  =A\left(  t\right)  +\theta^{\alpha}B_{\alpha}
\tag{17}%
\end{equation}

Notice that $\lambda$ takes the place of the cosmological constant and is
related with $b$ by $b=\frac{-\lambda}{2}$. The superfield solution (17) is
N=2 (chiral or antichiral two components spinors), has conformal symmetry in
the case $C=0$ and is not unique: as was pointed out in the references
[12,13,18] there exist a larger class of vacuum solutions. The dynamics of the
solution is very simple as is easy to see from eqs.(17), that is not the case
in the superspace $\left(  1,3\mid1\right)  $ as we showed in the previous Section.

With the description of both superspaces above, we pass now to compare them in
order to establish if a one to one mapping exists between these superspaces.
By simple inspection we can see that the fermionic part of the superspace
solutions (2) and (17) is mapped one to one, explicitly (for the $(1\mid2)$
superspace indexes 1 and 2 for $\alpha$ and $\beta$ are understood).%
\[
\nu_{\alpha}=-2\beta\overset{\circ}{\phi}_{\alpha}%
\]%
\[
2\sqrt{b}=\omega
\]%
\[
\theta^{1}\leftrightarrow\overline{\theta}^{\overset{.}{\alpha}},\ \theta
^{2}\leftrightarrow\theta^{\alpha}%
\]%
\[
\left(  \sigma^{0}\right)  _{\ \overset{.}{\alpha}}^{\alpha}\leftrightarrow
b_{1}^{2}%
\]
if the following conditions over the four dimensional solution hold%
\[
\alpha=\beta,\ \ \ \ \ \ Z_{\alpha}=\overline{Z}_{\overset{.}{\beta}}=0\ \
\]

For the bosonic part of the superfield solutions (17)\ and (4) no direct
relation exists between them. Only taking the limit of the constants
$\left\vert \mathbf{a}\right\vert \rightarrow\infty$ of the non-degenerate
superspace $\left(  1,3\mid1\right)  $ (i.e. going to the standard $\left(
1,3\mid1\right)  $ superspace) the Gaussian solution (7) goes to the same type
that the described in (17) for the $\left(  1\mid2\right)  $ superspace, with
$c_{1}\approx\sqrt{\lambda}$ and $c_{2}\approx\varphi_{0}$. And this fact is
non trivial: this happens because the chirality is explicitly restored in this
limit as we can easily seen from equations (3) when $\left\vert \mathbf{a}%
\right\vert \rightarrow\infty,\omega^{2}\rightarrow0$. It is clear that the
solution coming from four dimensional non-degenerate superspace is the
physical one because represents a semiclassical (Gaussian) state of the
Husimi's type [10,19]: and this is a \textit{direct consequence }of the
non-degenerate characteristic of the supermetric. The important role played by
the constants $\ a$ and $a^{\ast}$ in the extended line element (1) is
localize the physical state in a precise region of the space-time, as is
easily seen from expression (7). This fact can give some hints in order to
explain and to treat from the mathematical point of view the mechanism of
confinement, spontaneous compactification and other problems in high energy
particle physics that can have a topological origin [16].

\section{Supergravity as a gauge theory and the origin of the supermetric}

Now, we will give some answers and hints about the origin of the
non-degenerate supermetric under consideration and the structure of the
equations of motion derived the respective super-Lagrangian constructed from
it. The starting point is the \ $OSP(1/4)$ (some times called super-$SO(4,1)$)
superalgebra%
\[
\lbrack M_{AB},M_{CD}]=\epsilon_{C}\left(  _{A}M_{B}\right)  _{D}+\epsilon
_{D}\left(  _{A}M_{B}\right)  _{C}%
\]%
\begin{equation}
\lbrack M_{AB},Q_{C}]=\epsilon_{C}\left(  _{A}Q_{B}\right)
\ \ \ ,\ \ \ \left\{  Q_{A},Q_{B}\right\}  =2M_{AB} \tag{18}%
\end{equation}
here the indices $A,B,C...$ stay for $\alpha,\beta,\gamma...\left(
\overset{.}{\alpha},\overset{.}{\beta},\overset{.}{\gamma}...\right)
$spinorial indices:$\alpha,\beta\left(  \overset{.}{\alpha},\overset{.}{\beta
}\right)  =1,2\left(  \overset{.}{1},\overset{.}{2}\right)  .$ We define the
superconnection $A$ due the following "gauging"
\begin{equation}
A^{p}T_{p}\equiv\omega^{\alpha\overset{.}{\beta}}M_{\alpha\overset{.}{\beta}%
}+\omega^{\alpha\beta}M_{\alpha\beta}+\omega^{\overset{.}{\alpha}\overset
{.}{\beta}}M_{\overset{.}{\alpha}\overset{.}{\beta}}+\omega^{\alpha}Q_{\alpha
}-\omega^{\overset{.}{\alpha}}\overline{Q}_{\overset{.}{\alpha}} \tag{19}%
\end{equation}
where $\left(  \omega M\right)  $ define a ten dimensional bosonic
manifold$^{[3]}$\footnotetext[3]{Corresponding to the number of generators of
$SO\left(  4,1\right)  $ or $SO\left(  3,2\right)  $ that define the group
manifold.} and $p\equiv$multi-index, as usual. Analogically the
super-curvature is defined by $F\equiv F^{p}T_{p}$ with the following detailed
structure%
\begin{equation}
F\left(  M\right)  ^{AB}=d\omega^{AB}+\omega_{\ C}^{A}\wedge\omega^{CB}%
+\omega^{A}\wedge\omega^{B} \tag{20}%
\end{equation}%
\begin{equation}
F\left(  Q\right)  ^{A}=d\omega^{A}+\omega_{\ C}^{A}\wedge\omega^{C} \tag{21}%
\end{equation}
From (20-21) is not difficult to see, that there are a bosonic part and a
fermionic part associated with the even and odd generators of the
superalgebra. Our proposal for the action is%
\begin{equation}
S=\int F^{p}\wedge\mu_{p} \tag{22}%
\end{equation}
where the tensor $\mu_{p}$ (that play the role of a $OSP(1/4)$ diagonal
metric) is defined as%
\begin{equation}%
\begin{array}
[c]{ccc}%
\mu_{\alpha\overset{.}{\beta}}=\zeta_{\alpha}\wedge\overline{\zeta}%
_{\overset{.}{\beta}} & \mu_{\alpha\beta}=\zeta_{\alpha}\wedge\zeta_{\beta} &
\mu_{\alpha}=\nu\zeta_{\alpha}%
\end{array}
etc. \tag{23}%
\end{equation}
with $\zeta_{\alpha}\left(  \overline{\zeta}_{\overset{.}{\beta}}\right)  $
anticommuting spinors (suitable basis$^{[4]}$\footnotetext[4]{In general this
tensor has the same structure that the Cartan-Killing metric of the group
under consideration.}) and $\nu$ the parameter of the breaking of $OSP(1/4)$
to $SP(4)\sim$ $SO(4,1)$ symmetry of $\mu_{p}$.

In order to obtain the dynamical equations of the theory, we proceed to
perform the variation of the proposed action (22)%
\begin{align}
\delta S  &  =\int\delta F^{p}\wedge\mu_{p}+F^{p}\wedge\delta\mu_{p}\tag{24}\\
&  =\int d_{A}\mu_{p}\wedge\delta A^{p}+F^{p}\wedge\delta\mu_{p}\nonumber
\end{align}
where $d_{A}$ is the exterior derivative with respect to the $OSP(1/4)$
connection and: $\delta F=d_{A}\delta A$ have been used. Then, as a result,
the dynamics are described by
\begin{equation}
d_{A}\mu=0\ \ \ \ \ \ ,\ \ \ \ \ \ F=0\ \ \ \tag{25}%
\end{equation}
The fist equation said us that $\mu$ is covariantly constant with respect to
the $OSP(1/4)$ connection: this fact will be very important when the
$OSP(1/4)$ symmetry breaks down to $SP(4)\sim$ $SO(4,1)$ because $d_{A}%
\mu=d_{A}\mu_{AB}+d_{A}\mu_{A}=0,$ a soldering form will appear. The second
equation give the condition for a super Cartan connection $A=\omega
^{AB}+\omega^{A}$ to be flat, as is easily to see from the reductive
components of above expressions%
\begin{align}
F\left(  M\right)  ^{AB}  &  =R^{AB}+\omega^{A}\wedge\omega^{B}=0\tag{26}\\
F\left(  Q\right)  ^{A}  &  =d\omega^{A}+\omega_{\ C}^{A}\wedge\omega
^{C}=d_{\omega}\omega^{A}=0\nonumber
\end{align}
where now $d_{\omega}$ is the exterior derivative with respect to the
$SP(4)\left(  \sim SO(4,1)\right)  $ connection and $R^{AB}\equiv d\omega
^{AB}+\omega_{\ C}^{A}\wedge\omega^{CB}\ $is the\ $SP(4)\sim$ $SO(4,1)$%
\ curvature. Then%
\begin{equation}
F=0\Leftrightarrow R^{AB}+\omega^{A}\wedge\omega^{B}=0\text{ and }d_{\omega
}\omega^{A}=0 \tag{27}%
\end{equation}
the second condition says that the $SP(4)\left(  \sim SO(4,1)\right)  $
connection is super-torsion free. The first says not that the $SP(4)\left(
\sim SO(4,1)\right)  $ connection is flat but that it is homogeneous with a
cosmological constant\ related to the explicit structure of the Cartan forms
$\omega^{A}$, as we will see when the $OSP(1/4)$ action is reduced to the
Volkov-Pashnev model.

\subsection{The geometrical reduction: origin of the extended supermetric}

The supermetric under consideration, proposed by Volkov and Pashnev in [9],
can be obtained from the $OSP(1/4)$ action via the following procedure:

i) the In\"{o}nu-Wigner contraction [32] in order to pass from $OSP(1/4)$ to
the super-Poincare algebra (corresponding to the original symmetry of the
model of refs.[10,9]), then, the \textit{even} part of the curvature is
splitted into a $\mathbb{R}^{3,1}$ part $R^{\alpha\overset{.}{\beta}}$ and a
$SO\left(  3,1\right)  $ part $R^{\alpha\beta}\left(  R^{\overset{.}{\alpha
}\overset{.}{\beta}}\right)  $associated with the remaining six generators of
the original $SP(4)$ group. This fact is easily realized knowing that the
underlying geometry is reductive: $SP(4)\sim SO\left(  4,1\right)  \rightarrow
SO\left(  3,1\right)  +$ $\mathbb{R}^{3,1}$, and rewriting the superalgebra
(18) as%
\begin{equation}%
\begin{array}
[c]{ccc}%
\lbrack M,M]\sim M & [M,\Pi]\sim\Pi & [\Pi,\Pi]\sim M\\
\lbrack M,S]\sim S & [\Pi,S]\sim S & \left\{  S,S\right\}  \sim M+\Pi
\end{array}
\tag{28}%
\end{equation}
$\left(  \text{with }\Pi\sim M_{\alpha\overset{.}{\beta}}\text{ and }M\sim
M_{\alpha\beta}\left(  M_{\overset{.}{\alpha}\overset{.}{\beta}}\right)
\right)  $ and rescales $m^{2}\Pi=P$ and $mS=Q$, in the limit $m\rightarrow0$
one recovers the super Poincare algebra. Notice that one does not rescale $M$
since one want to keep $[M,M]\sim M$ Lorentz algebra (that also is symmetry of (1))

ii)the spontaneous break down of the $OSP(1/4)$ to the $SP(4)\sim SO\left(
4,1\right)  $ symmetry of $\mu_{p}$(e.g: $\nu\rightarrow0$ in $\mu_{p}$) of
such manner that the even part of the $OSP(1/4)$ action $F\left(  M\right)
^{AB}$ remains. (Notice the important fact that the super-action under
consideration \textit{carry naturally} the fermionic "Dirac type" part that
disappears with the particular breakdown of the symmetry of $\mu_{p}$, this
fact was not having account in several refs. as in [38])

\textit{After} these processes have been explicitly realized, the
\textit{even} part of the original $OSP(1/4)$ action (now super-Poincare
invariant ) can be related with the original metric (1) as follows:
\begin{equation}
R\left(  M\right)  +R\left(  P\right)  +\omega^{\alpha}\omega_{\alpha}%
-\omega^{\overset{.}{\alpha}}\omega_{\overset{.}{\alpha}}\rightarrow
\omega^{\mu}\omega_{\mu}+\mathbf{a}\omega^{\alpha}\omega_{\alpha}%
-\mathbf{a}^{\ast}\omega^{\overset{.}{\alpha}}\omega_{\overset{.}{\alpha}}%
\mid_{VP} \tag{29}%
\end{equation}
Notice that there is mapping $R\left(  M\right)  +R\left(  P\right)
\rightarrow\omega^{\mu}\omega_{\mu}\mid_{VP}$ that is well defined and can be
realized of different forms, and the map of interest here $\omega^{\alpha
}\omega_{\alpha}-\omega^{\overset{.}{\alpha}}\omega_{\overset{.}{\alpha}%
}\rightarrow\mathbf{a}\omega^{\alpha}\omega_{\alpha}-\mathbf{a}^{\ast}%
\omega^{\overset{.}{\alpha}}\omega_{\overset{.}{\alpha}}\mid_{VP}$ that
associate the Cartan forms of the original $OSP(1/4)$action (22) with the
Cartan forms of the Volkov-Pashnev supermodel: $\omega^{\alpha}=\left(
\mathbf{a}\right)  ^{1/2}\omega^{\alpha}\mid_{VP};$ $\omega^{\overset
{.}{\alpha}}=\left(  \mathbf{a}^{\ast}\right)  ^{1/2}\omega^{\overset
{.}{\alpha}}\mid_{VP}.$ Then, the origin of the coefficients $\mathbf{a}$ and
$\mathbf{a}^{\ast}$becomes clear from the geometrical point of view.

What about physics? From the first condition in (27) and the association (29)
it is not difficult to see that, as in the case of the spacetime cosmological
constant $\Lambda:R=\frac{\Lambda}{3}e\wedge e\ \left(  e\equiv
space-time\ tetrad\right)  $, there is a cosmological term from the superspace
related to the complex parameters $\mathbf{a}$ and $\mathbf{a}^{\ast}$ :
$R=-\left(  \mathbf{a}\omega^{\alpha}\omega_{\alpha}-\mathbf{a}^{\ast}%
\omega^{\overset{.}{\alpha}}\omega_{\overset{.}{\alpha}}\right)  $ and is
easily to see from the minus sign in above expression, why for the standard
supersymmetric\ (supergravity) models is more natural to use $SO\left(
3,2\right)  $ instead $SO\left(  4,1\right)  ..$

On the associated spinorial action in the action (22), notice that the role of
this part is constrained by the nature of $\nu\zeta_{\alpha}$ in $\mu_{p}$:

i) If they are of the same nature of the $\omega^{\alpha}$ this term is a
total derivative, has not influence into the equations of motion, then the
action proposed by Volkov and Pashnev in [9,10] has the correct fermionic form.

ii) If they are not with the same $SP(4)\sim SO\left(  4,1\right)  $
invariance that the $\omega^{\alpha}$, the symmetry of the original model has
been modified. In this direction a relativistic supersymmetric model for
particles was proposed in ref. [33] considering an N-extended Minkowski
superspace and introducing central charges to the superalgebra. Hence the
underlying rigid symmetry gets enlarged to N-extended super-Poincare algebra.
Considering for our case similar superextension that in ref.[33] we can
introduce the following new action
\begin{equation}
S=-m\int_{\tau1}^{\tau2}d\tau\sqrt{\overset{\circ}{\omega_{\mu}}\overset
{\circ}{\omega^{\mu}}+a\overset{.}{\theta}^{\alpha}\overset{.}{\theta}%
_{\alpha}-a^{\ast}\overset{.}{\overline{\theta}}^{\overset{.}{\alpha}}%
\overset{.}{\overline{\theta}}_{\overset{.}{\alpha}}+i(\theta^{\alpha i}%
A_{ij}\overset{.}{\theta}_{\alpha}^{j}-\overline{\theta}^{\overset{.}{\alpha
}i}A_{ij}\overset{.}{\overline{\theta}}_{\overset{.}{\alpha}}^{j})}=\int
_{\tau1}^{\tau2}d\tau L\left(  x,\theta,\overline{\theta}\right)  \tag{30}%
\end{equation}
that is the N-extended version of the superparticle model proposed in [9],
with the first order fermionic part. The matrix tensor $A_{ij}$ introduce the
simplectic structure of such manner that now $\zeta_{\alpha i}\sim
A_{ij}\theta_{\alpha}^{j}$ is not covariantly constant under $d_{\omega}%
.\ $Notice that the "Dirac-like" fermionic part is obviously \textit{inside
}the square root because it is part of the full curvature (the geometry of a
N-superspace), fact that was not advertised by the authors in [33] that,
specifically in they work, they not take account on the geometrical origin of
the action.\textit{ }The interesting point is perform the same quantization
that in the first part of this research [10] in order to obtain and compare
the spectrum of physical states with the obtained in ref.[33]. This issue will
be presented elsewhere [16].

In resume, we explicitly show here that the action under consideration
constructed with the non-degenerate supermetric, can be derived from a
$OSP(1/4)$ action and the structure of the dynamical equations for the fields
of the theory depends on the coefficients of the tensor $\mu_{p}$ because they
are responsible of the manner that the $OSP(1/4)$ symmetry of the model is
breakdown, and a new N-extended version of the supermodel of [9] is presented.

\section{ "Warped" gravity models, confinement and the supermetric}

It is well known that large extra dimensions offer an opportunity for a new
solution to the hierarchy problem [20]. Field theoretical localization
mechanisms for scalar and fermions [21] as well as for gauge bosons [22] were
found. The crucial ingredient of this scenario is a brane on which standard
model particles are localized. In string theory, fields can naturally be
localized on D-branes due to the open strings ending on them[23]. Up until
recently, extra dimensions had to be compactified, since the localization
mechanism for gravity was not known. It was suggested in ref.[24] that
gravitational interactions between particles on a brane in uncompactified five
dimensional space could have the correct four dimensional Newtonian behaviour,
provided that the bulk cosmological constant and the brane tension are
related. Recently, it was found by Randall and Sundrum that gravitons can be
localized on a brane which separates\ two patches of AdS$_{5}$ space-time
[25]. The necessary requirement for the four-dimensional brane Universe to be
static is that the tension of the brane is fine-tuned to the bulk cosmological
constant [24,25]. By the other hand, recent papers present an interesting
model in which the extra dimensions are used only as a mathematical tool
taking advantage of the AdS/CFT correspondence that claims that the 5D warped
dimension is related with a strongly coupled 4D theory [26].

A remarkable property of the solution given by the expression (7) is that the
physical state $g_{ab}\left(  x\right)  $ is localized in a particular
position of the space-time. The supermetric coefficients $\boldsymbol{a}$ and
$\boldsymbol{a}^{\ast}$play the important role of localize the fields in the
bosonic part of the superspace in similar and suggestive form as the well
known "warp factors" in multidimensional gravity[14] for a positive (or
negative) tension brane. But the essential difference is, because the
$\mathbb{C}$-constants $\mathbf{a}$ and $\mathbf{a}^{\ast}$ coming from the
$B_{L,0}$(even) fermionic part of the superspace under consideration, not
additional and/or topological structures that break the symmetries of the
model (i.e. reflection $Z_{2}$-symmetry) are required: the natural structure
of the superspace produces this effect.

Also it is interesting to remark here that the Gaussian type solution (7) is
very well defined physical state in a Hilbert space[10,19] from the
mathematical point of view, contrarily to the case $u\left(  y\right)
=ce^{-H\left\vert y\right\vert }$ given in [14] that, although were possible
to find a manner to include it in any Hilbert space, is strongly needed to
take special mathematical and physical particular assumptions whose meaning is
obscure. The comparison with the case of 5-dimensional gravity plus
cosmological constant[14] is given in the following table:%
\[%
\begin{tabular}
[c]{|l|l|l|}\hline
$Spacetime$ & $\left(  5-D\right)  \text{ }gravity+\Lambda$ &
$Superspace\left(  1,d\mid1\right)  $\\\hline
$Interval$ & $ds^{2}=A\left(  y\right)  dx_{3+1}^{2}-dy$ & $ds^{2}=\omega
^{\mu}\omega_{\mu}+\mathbf{a}\omega^{\alpha}\omega_{\alpha}-\mathbf{a}^{\ast
}\omega^{\overset{.}{\alpha}}\omega_{\overset{.}{\alpha}}$\\\hline
$Equation$ & $\left[  -\partial_{y}^{2}-m^{2}e^{H\left\vert y\right\vert
}+H^{2}-2H\delta\left(  y\right)  \right]  u\left(  y\right)  =0$ & $\left[
\left\vert a\right\vert ^{2}\left(  \partial_{0}^{2}-\partial_{i}^{2}\right)
+\frac{1}{4}\left(  \partial_{\eta}-\partial_{\xi}+i\ \partial_{\mu}\left(
\sigma^{\mu}\right)  \xi\right)  ^{2}-\right.  $\\
&  & $\left.  -\frac{1}{4}\left(  \partial_{\eta}+\partial_{\xi}%
+i\ \partial_{\mu}\left(  \sigma^{\mu}\right)  \xi\right)  ^{2}+m^{2}\right]
_{cd}^{ab}g_{ab}=0$\\\hline
$Solution$ & $u\left(  y\right)  =ce^{-H\left\vert y\right\vert }%
,\ \ \ H\equiv\sqrt{-\frac{2\Lambda}{3}}=\frac{\left\vert T\right\vert }%
{M^{3}}$ & $g_{ab}\left(  x\right)  =e^{-\left(  \frac{m}{\left\vert
\mathbf{a}\right\vert }\right)  ^{2}x^{2}+c_{1}^{\prime}x+c_{2}^{\prime}%
}e^{\xi\varrho\left(  x\right)  }\left\vert f\left(  \xi\right)  \right\vert
^{2}\left(
\begin{array}
[c]{c}%
\alpha\\
\alpha^{\ast}%
\end{array}
\right)  _{ab}$\\\hline
\end{tabular}
\ \ \ \ \ \ \ \ \ \ \ \ \ \
\]
Here, in order to make our comparison consistent, the proposed superspace has
$d=n+4$ bosonic coordinates and the extended superspace solution for $n=0$ can
depend, in principle, on any or all the 4-dimensional coordinates:
$x\equiv\left(  t,\overline{x}\right)  $, $c_{1}^{\prime}x\equiv c_{1\mu
}^{\prime}x^{\mu}$ and $c_{2}^{\prime}$ scalar (e.g.: the $t$ coordinate in
expression (7)); for $n\neq0$ it depends on the $n$-additional coordinates.

Notice the following important observations:

i) that for that the solution in the 5-dimensional gravity plus $\Lambda$
case, the explicit presence of the cosmological term is necessary for the
consistency of the model: the "fine-tuning" $H\equiv\sqrt{-\frac{2\Lambda}{3}%
}=\frac{\left\vert T\right\vert }{M^{3}}$, where $T$ is the tension of the
brane and $M^{3}$ is the constant of the Einstein-Hilbert +$\Lambda$ action.

ii) about the localization of the fields given by the particular superspace
treated here, the $Z_{2}$ symmetry is non-compatible with the solution that
clearly is not chiral or antichiral. This fact is consistent with the analysis
given for a similar superspace that the considered here in ref.[10,27] where
the solutions are superprojected in a sector of the physical states that is
not chiral or antichiral.

iii) because for $n=0$ our solution (7) is attached on the 3+1 space-time but
the localization occurs on the time coordinate (on in any of the remanent 3
space coordinates) the physics seems to be very different with respect to the
warped gravity model where the field equation in final form for the
5-dimensional gravity depends on the extra dimension$^{[5]}\footnotetext[5]%
{e.g.: in the Randall-Sundrum model the graviton is localized in the extra
dimension}.$ This $n=0$ \ case can give some hints for the\ theoretical
treatment of the confinement mechanism with natural breaking of the chiral
symmetry in high energy physics (e.g.:instanton liquid models, etc);

iv) for $n=1$, our model with the solution depending on the extra coordinate,
the situation changes favorably: the localization of the field is in the
additional bosonic coordinate (as the graviton in the RS type model) but with
all the good properties of the solution (7) already mentioned in the beginning
of this paragraph.

From the points discussed above and the "state of the art " of the problem, we
seen the importance of to propose new mechanisms and alternative models that
can help us to understand and to handle the problem. Also is clearly important
that the supermetric (1), cornerstone of this simple supermodel, is
non-degenerate in order to solve in a simultaneous manner the
localization-confinement of the fields involved and the breaking of the chiral
symmetry. Then, it is not difficult to think to promote the particular
supermetric under study towards to build a strongly coupled 4D model, using
this particular N=1 toy superspace. We will treat this issue with great detail
in a further work[16].

\section{Generalized phase-space representations: "tomographic" interpretation
of the physic states}

In this section we will treat to elucidate the meaning of the basic
(non-observable) states and the physical ones of Section 2: $\Psi$ and
$g_{ab}$ respectively. To this end, before to enter in more technical
questions, some important points of the toy model presented here need to be
reminded from the previous Sections and from reference [10] (mainly in Section 5):

$\circ$ For simplicity, only the bosonic part of the superfield wave solution
will be analyzed in order to discuss conveniently the meaning of the states
involved: the fermionic part, e.g.: $e^{\xi\varrho\left(  \alpha+\alpha^{\ast
}\right)  }\left\vert f\left(  \xi\right)  \right\vert ^{2}$ will be not
discussed here.

$\circ$ The detailed mathematical structure of the basic states $\Psi
_{3/4}\left(  t,\xi,q\right)  $ and $\Psi_{1/4}\left(  t,\xi,q\right)  $ will
be not considered (they will be studied elsewhere [16]).

$\circ$There are two spaces, that we denote $\mathcal{H}$ and $\overline
{\mathcal{H}}$ (dotted and undotted indices, as usual for different helicity
states), each one being the direct sum of two irreducible subspaces
$\mathcal{H=H}_{1/4}\oplus\mathcal{H}_{3/4}$ $\left(  \overline{\mathcal{H}%
}\mathcal{=}\overline{\mathcal{H}}_{1/4}\oplus\overline{\mathcal{H}}%
_{3/4}\right)  $

$\circ$Each irreducible subspace $\mathcal{H}_{1/4}$ (\text{with spin
1/4)}$,\mathcal{H}_{3/4}$ (with spin 3/4) are spanned by even and odd states
states $\left\vert 2n\right\rangle $ and $\left\vert 2n+1\right\rangle $
respectively. They are eigenstates of $N=a^{+}a\sim K_{0}$, that in more
standard form:%
\begin{align*}
\left\vert n\right\rangle  &  =\left\vert \frac{1}{4},\frac{1}{2}\left(
n+\frac{1}{2}\right)  \right\rangle \text{ \ \ \ \ \ \ \ \ \ for n even}\\
\left\vert n\right\rangle  &  =\left\vert \frac{3}{4},\frac{1}{2}\left(
n+\frac{1}{2}\right)  \right\rangle \text{\ \ \ \ \ \ \ \ \ \ for n odd}%
\end{align*}

$\circ$ The specific \textit{basic} solutions naturally obtained from the
non-degenerate superspace (see Section 2) are coherent states $\Psi
_{1/4}\left(  t,\xi,q\right)  $ and $\Psi_{3/4}\left(  t,\xi,q\right)  ,$
eigenstates of the operator $K_{-}=\frac{1}{2}aa,\ $spanning respectively the
irreducible subspaces $\mathcal{H}_{1/4}$ and $\mathcal{H}_{3/4}$(also for the
"dotted" case $\overline{\mathcal{H}}_{1/4}$ and $\overline{\mathcal{H}}%
_{3/4}$).

$\circ$ $\Psi_{1/4}\left(  t,\xi,q\right)  $ and $\Psi_{3/4}\left(
t,\xi,q\right)  $ are mutually orthogonal ($\left\langle \Psi_{3/4}\right.
\left\vert \Psi_{1/4}\right\rangle =\left\langle \Psi_{1/4}\right.  \left\vert
\Psi_{3/4}\right\rangle =0$).

$\circ$ One general state of any spin can be expanded in the $\left\vert
n\right\rangle $ or in the CS $\left\vert \Psi\right\rangle $ basis , due the
well known properties of such states.

$\circ$ From the previous points and the explicit solution of the (super)
relativistic wave equation, notice that there are four (non-trivial)
representations for the group decomposition of the bispinor solution$,$as
follows: $\left(  1/4,0\right)  \oplus\left(  0,3/4\right)  ,$ $\left(
3/4,0\right)  \oplus\left(  0,1/4\right)  ,\left(  1/4,0\right)  \oplus\left(
0,1/4\right)  $ and $\left(  3/4,0\right)  \oplus\left(  0,3/4\right)  $
(Section 5 ref.[10])$.$

Is very well known, the quality of the CS of being "canonical quantizers"
[37]. The CS quantization (BKT, Berezin-Klauder-Toeplitz) consists in
associating with any classical observable $f$ (function of the phase space
variables $q,p$ or equivalently $z,\overline{z}$) the operator valuated
integral%
\[
\frac{1}{\pi}\int_{\mathbb{C}}f\left(  z,\overline{z}\right)  \left\vert
z\right\rangle \left\langle z\right\vert dz^{2}=A_{f}%
\]
The resulting operator (if it exists, almost in a weak sense), acts on the
Hilbert space, spanned by the (over) complete set of CS $\left\vert
z\right\rangle $. In the weak sense we mean that the integral%
\[
\frac{1}{\pi}\int_{\mathbb{C}}f\left(  z,\overline{z}\right)  \left\vert
\left\langle \varphi\right.  \left\vert z\right\rangle \right\vert ^{2}%
dz^{2}=\left\langle \varphi\left\vert A_{f}\right\vert \varphi\right\rangle
\]
should be finite by any $\left\vert \varphi\right\rangle \in\mathcal{H}$ (or
$\in$ to some dense subset in $\mathcal{H}$). On immediately notices that if
$\varphi$ is normalized then the previous equation represents the mean value
of the $f$ function with respect to the $\varphi$-dependent probability
distribution $\left\vert \left\langle \varphi\right.  \left\vert
z\right\rangle \right\vert ^{2}$on the phase space. Because this can be
thought as a de-quantization, if we take account on one of the fundamental
features of the CS that is the resolution of the unity in any Hilbert space
$\mathcal{H}$, the bridge between the classical and the quantum world can be
established. Then, the fact that the obtained basic states of the superfield
solution presented here are CS states is clearly important from the classical
and quantum point of view.

We will see soon that there exist some type of operators, $L_{ab}$ for
example$,$ where the integral as in the above equation involves
\textit{non-diagonal} projectors. That means that is necessary an extension of
the set of "acceptable classical observables" to those CS distributions $T$
$\in$ $\mathcal{D}^{\prime}\left(  \mathbb{R}^{2}\right)  $ (space of
distributions) such that the product $e^{-\eta\left\vert z\right\vert ^{2}}T$
$\in$ $\mathcal{S}^{\prime}\left(  \mathbb{R}^{2}\right)  ,$( e.g. tempered
distributions in $\mathcal{D}^{\prime}\left(  \mathbb{R}^{2}\right)  $that
belong to the Schwartz space $\mathcal{S}^{\prime}\left(  \mathbb{R}%
^{2}\right)  $). As was pointed out in ref.[35], an increase in the family of
representations of various systems offers new ways to study such systems.
Representations of Hilbert space operators in the manner of the Weyl
representation may be carried out for a great variety of groups, asymmetric
representations of various forms (analogous to those presented in [35] for the
Weyl group) can be introduced for other groups. In our case, the big group
involved is the metaplectic group $Mp\left(  2\right)  $ (the covering group
of $SL(2C)$). This important group $Mp\left(  2\right)  $ have been studied
with some detail in several references [36] and is closely related with the
para-bose coherent and squeezed states (CS\ and SS).

The starting point for our analysis is the following CS\ reconstructing Kernel
for any operator A (not necessarily bounded[37,19])%

\begin{equation}
K_{\widehat{A}}\left(  \alpha,\alpha^{\prime};g\right)  =e^{\left[  \left\vert
\alpha\right\vert ^{2}-\left\vert \alpha^{\prime}\right\vert ^{2}\right]
}\left\langle \alpha\left\vert A\right\vert \alpha^{\prime}\right\rangle
\tag{31}%
\end{equation}
where $\alpha$ and $\alpha^{\prime}$are complex variables that characterize a
respective CS (in principle can depend on time, see expressions (32) below)
and $g$ is an element of $Mp\left(  2\right)  $. From the first paper of this
work and the previous sections, the possible \textit{basic states} (CS not
physically observables, regard the discussion in Section 2) are classified as%
\[
\left\vert \Psi_{1/4}\left(  t,\xi,q\right)  \right\rangle =f\left(
\xi\right)  \left\vert \alpha_{+}\left(  t\right)  \right\rangle
\]%
\begin{equation}
\left\vert \Psi_{3/4}\left(  t,\xi,q\right)  \right\rangle =f\left(
\xi\right)  \left\vert \alpha_{-}\left(  t\right)  \right\rangle \tag{32a}%
\end{equation}
with the following symmetric and antisymmetric non-equivalent combinations%
\[
\left\vert \Psi^{S}\right\rangle =\frac{f\left(  \xi\right)  }{\sqrt{2}%
}\left(  \left\vert \alpha_{+}\right\rangle +\left\vert \alpha_{-}%
\right\rangle \right)  =f\left(  \xi\right)  \left\vert \alpha^{S}\left(
t\right)  \right\rangle
\]%
\begin{equation}
\left\vert \Psi^{A}\right\rangle =\frac{f\left(  \xi\right)  }{\sqrt{2}%
}\left(  \left\vert \alpha_{+}\right\rangle -\left\vert \alpha_{-}%
\right\rangle \right)  =f\left(  \xi\right)  \left\vert \alpha^{A}\left(
t\right)  \right\rangle \tag{32b}%
\end{equation}
and the important fact in order to evaluate the kernels (31) was the action of
$a$ and $a^{2}$ over the states previously defined%
\[
a\left\vert \Psi_{1/4}\right\rangle =\alpha\left\vert \Psi_{3/4}\right\rangle
;\ a\left\vert \Psi_{3/4}\right\rangle =\alpha\left\vert \Psi_{1/4}%
\right\rangle ;\ a\left\vert \Psi^{S}\right\rangle =\alpha\left\vert \Psi
^{S}\right\rangle ;\ a\left\vert \Psi^{A}\right\rangle =-\alpha\left\vert
\Psi^{A}\right\rangle
\]%
\[
a^{2}\left\vert \Psi_{1/4}\right\rangle =\alpha^{2}\left\vert \Psi
_{1/4}\right\rangle ;\ a^{2}\left\vert \Psi_{3/4}\right\rangle =\alpha
^{2}\left\vert \Psi_{3/4}\right\rangle ;\ a^{2}\left\vert \Psi^{S}%
\right\rangle =\alpha^{2}\left\vert \Psi^{S}\right\rangle ;\ a^{2}\left\vert
\Psi^{A}\right\rangle =\alpha^{2}\left\vert \Psi^{A}\right\rangle
\]
(similarly for states $\overline{\Psi}$ ). We have that the \textit{physical
states} are particular representations of the operators $L_{ab}$ and
$\mathbb{L}_{ab}$ $\left(  \in Mp\left(  2\right)  \right)  $in spinorial form
in the sense of quasiprobabilities (tomograms in the $\Psi_{s}$ plane) or as
mean values with respect to the basic CS (32): $\left\vert \Psi_{\lambda
}\right\rangle \left(  \lambda=1/4,1/2,3/4,1\right)  $. The possible
generalized kernels(31) are the following
\[
\left.  g_{ab}\left(  t,2,\alpha\right)  \right\vert _{hw}=\left\langle
\Psi^{S}\left(  t\right)  \right\vert L_{ab}\left\vert \Psi^{S}\left(
t\right)  \right\rangle =e^{-\left(  \frac{m}{\sqrt{2}\left\vert
\mathbf{a}\right\vert }\right)  ^{2}\left[  \left(  \alpha+\alpha^{\ast
}\right)  -B\right]  ^{2}+D}e^{\xi\varrho\left(  \alpha+\alpha^{\ast}\right)
}\left\vert f\left(  \xi\right)  \right\vert ^{2}\left(
\begin{array}
[c]{c}%
\alpha\\
\alpha^{\ast}%
\end{array}
\right)  _{\left(  2\right)  ab}%
\]%
\[
\left.  g_{ab}\left(  t,1,-\alpha\right)  \right\vert _{hw}=\left\langle
\Psi^{A}\left(  t\right)  \right\vert L_{ab}\left\vert \Psi^{A}\left(
t\right)  \right\rangle =e^{-\left(  \frac{m}{\sqrt{2}\left\vert
\mathbf{a}\right\vert }\right)  ^{2}\left[  \left(  \alpha+\alpha^{\ast
}\right)  -B\right]  ^{2}+D}e^{\xi\varrho\left(  \alpha+\alpha^{\ast}\right)
}\left\vert f\left(  \xi\right)  \right\vert ^{2}\left(
\begin{array}
[c]{c}%
-\alpha\\
-\alpha^{\ast}%
\end{array}
\right)  _{\left(  1\right)  ab}%
\]%
\[
\left.  g_{ab}\left(  t,2,\alpha^{2}\right)  \right\vert _{su(1,1)}%
=\left\langle \Psi^{S}\left(  t\right)  \right\vert \mathbb{L}_{ab}\left\vert
\Psi^{S}\left(  t\right)  \right\rangle =e^{-\left(  \frac{m}{\sqrt
{2}\left\vert \mathbf{a}\right\vert }\right)  ^{2}\left[  \left(
\alpha+\alpha^{\ast}\right)  -B\right]  ^{2}+D}e^{\xi\varrho\left(
\alpha+\alpha^{\ast}\right)  }\left\vert f\left(  \xi\right)  \right\vert
^{2}\left(
\begin{array}
[c]{c}%
\alpha^{2}\\
\alpha^{\ast2}%
\end{array}
\right)  _{\left(  2\right)  ab}%
\]%
\[
\left.  g_{ab}\left(  t,1,\alpha^{2}\right)  \right\vert _{su(1,1)}%
=\left\langle \Psi^{A}\left(  t\right)  \right\vert \mathbb{L}_{ab}\left\vert
\Psi^{A}\left(  t\right)  \right\rangle =e^{-\left(  \frac{m}{\sqrt
{2}\left\vert \mathbf{a}\right\vert }\right)  ^{2}\left[  \left(
\alpha+\alpha^{\ast}\right)  -B\right]  ^{2}+D}e^{\xi\varrho\left(
\alpha+\alpha^{\ast}\right)  }\left\vert f\left(  \xi\right)  \right\vert
^{2}\left(
\begin{array}
[c]{c}%
\alpha^{2}\\
\alpha^{\ast2}%
\end{array}
\right)  _{\left(  1\right)  ab}%
\]%
\[
\left.  g_{ab}\left(  t,3/2,\alpha^{2}\right)  \right\vert _{su(1,1)}%
=\left\langle \Psi_{3/4}\left(  t\right)  \right\vert \mathbb{L}%
_{ab}\left\vert \Psi_{3/4}\left(  t\right)  \right\rangle =e^{-\left(
\frac{m}{\sqrt{2}\left\vert \mathbf{a}\right\vert }\right)  ^{2}\left[
\left(  \alpha+\alpha^{\ast}\right)  -B\right]  ^{2}+D}e^{\xi\varrho\left(
\alpha+\alpha^{\ast}\right)  }\left\vert f\left(  \xi\right)  \right\vert
^{2}\left(
\begin{array}
[c]{c}%
\alpha^{2}\\
\alpha^{\ast2}%
\end{array}
\right)  _{(3/2)ab}%
\]%
\begin{equation}
\left.  g_{ab}\left(  t,1/2,\alpha^{2}\right)  \right\vert _{su(1,1)}%
=\left\langle \Psi_{1/4}\left(  t\right)  \right\vert \mathbb{L}%
_{ab}\left\vert \Psi_{1/4}\left(  t\right)  \right\rangle =e^{-\left(
\frac{m}{\sqrt{2}\left\vert \mathbf{a}\right\vert }\right)  ^{2}\left[
\left(  \alpha+\alpha^{\ast}\right)  -B\right]  ^{2}+D}e^{\xi\varrho\left(
\alpha+\alpha^{\ast}\right)  }\left\vert f\left(  \xi\right)  \right\vert
^{2}\left(
\begin{array}
[c]{c}%
\alpha^{2}\\
\alpha^{\ast2}%
\end{array}
\right)  _{(1/2)ab} \tag{33}%
\end{equation}
where the constants $D$ and $B$ are related to the original constants
$c_{1}^{\prime}$ and $c_{2}^{\prime}$ of the first part of this work as:
$D=\left(  \frac{\left\vert \mathbf{a}\right\vert c_{1}^{\prime}}{\sqrt{2}%
m}\right)  ^{2}+c_{2}^{\prime}$ and $B=\left(  \frac{\left\vert \mathbf{a}%
\right\vert }{m}\right)  ^{2}c_{1}^{\prime}.$The expressions (33) are in the
called Sudarshan's diagonal-representation that lead, as important
consequence, the \textit{physical states} with spin content $\lambda
=1/2,1,3/2,2$. Precisely, the CS generate a map that relates the solution of
the wave equation $g_{ab}$ to the specific subspace of the full Hilbert space
where these CS\ live (see the reconstruction of the operators $L$ and
$\mathbb{L}$ eqs. (35-37) below$\mathbb{)}$.

However, there exists for operators $\in Mp\left(  2\right)  $ an asymmetric
-kernel leading for our case the following $\lambda=1$ state%
\begin{align}
\left.  g_{ab}\left(  t,1,\alpha\right)  \right\vert _{hw}  &  =\left\langle
\Psi_{3/4}\left(  t\right)  \right\vert L_{ab}\left\vert \Psi_{1/4}\left(
t\right)  \right\rangle =\tag{34}\\
&  =\left\langle \Psi_{1/4}\left(  t\right)  \right\vert L_{ab}\left\vert
\Psi_{3/4}\left(  t\right)  \right\rangle =e^{-\left(  \frac{m}{\sqrt
{2}\left\vert \mathbf{a}\right\vert }\right)  ^{2}\left[  \left(
\alpha+\alpha^{\ast}\right)  -B\right]  ^{2}+D}e^{\xi\varrho\left(
\alpha+\alpha^{\ast}\right)  }\left\vert f\left(  \xi\right)  \right\vert
^{2}\left(
\begin{array}
[c]{c}%
\alpha\\
\alpha^{\ast}%
\end{array}
\right)  _{\left(  1\right)  ab}\nonumber
\end{align}
because the non-diagonal projector involved in the reconstruction formula of
$L_{ab}$ is formed with $\Psi_{1/4}$ and $\Psi_{3/4}$ spanning all the Hilbert space.

Observation 1: Due the unobservability of isolated basic states, the spin zero
physical states appears as bounded states $g\overline{g}$ , where
$g_{ab}\left(  t,s,w\right)  $ and $\overline{g}_{ab}\left(  t,s,w\right)  $
are given by the bilinear expressions (33).

Observation 2: Each kernel represents a \textit{physical} state composed by
fundamental ones, that are \textit{basic} and \textit{unobservable}.

Notice that the spectrum of the physic states are labeled not only by they
spin content $\lambda$, but also for the "eigenspinors\footnotetext[6]{This
term was introduced here by us.}"$^{\text{[6]}}\left(
\begin{array}
[c]{c}%
\alpha\\
\alpha^{\ast}%
\end{array}
\right)  _{\left(  \lambda\right)  ab}$corresponding to the tomographic
representations of $L_{ab}$(map over a region of $\mathcal{H}$) $;$ and
$\left(
\begin{array}
[c]{c}%
\alpha^{2}\\
\alpha^{\ast2}%
\end{array}
\right)  _{(\lambda)ab}$corresponding to the tomographic representations of
$\mathbb{L}_{ab}$ .

The previous representations form part of a more general class of
representations for operators recently proposed by Klauder and Skagerstam in
ref.[35]. However, the operators can be reconstructed due the (over)
completeness of the basic states $\left\vert \Psi_{1/4}\left(  t\right)
\right\rangle $ in $\mathcal{H}_{1/4},$ $\left\vert \Psi_{3/4}\left(
t\right)  \right\rangle $ in $\mathcal{H}_{3/4}$ \ and $\left\vert \Psi
^{S}\left(  t\right)  \right\rangle $ $\left(  \left\vert \Psi^{A}\left(
t\right)  \right\rangle \right)  $in the full Hilbert space $\mathcal{H=H}%
_{1/4}\oplus\mathcal{H}_{3/4}$ through the following reconstruction formulas
(analogically for the states $\overline{\Psi})$:%
\begin{align}
\mathbb{L}_{ab}  &  =\int\frac{d^{2}\alpha}{\pi}\left[  \int\frac{d^{2}w}{\pi
}\int\frac{d^{2}\alpha^{\prime}}{\pi}e^{-\left(  \frac{m}{\sqrt{2}\left\vert
\mathbf{a}\right\vert }\right)  ^{2}\left[  \left(  \alpha+\alpha^{\ast
}\right)  -B\right]  ^{2}+D}e^{\xi\varrho\left(  \alpha^{\prime}%
+\alpha^{\prime\ast}\right)  }\left\vert f\left(  \xi\right)  \right\vert
^{2}\left(
\begin{array}
[c]{c}%
\alpha^{\prime2}\\
\alpha^{\ast\prime2}%
\end{array}
\right)  _{\left(  2\right)  ab}\right.  \times\tag{35a}\\
&  \times\left.  e^{\frac{\left\vert w\right\vert ^{2}}{4}}e^{\frac{i}%
{2}\left[  \left(  \alpha-\alpha^{\prime}\right)  w^{\ast}+\left(
\alpha^{\ast}-\alpha^{\ast\prime}\right)  w\right]  }\right]  \left\vert
\Psi^{S}\left(  \alpha\right)  \right\rangle \left\langle \Psi^{S}\left(
\alpha\right)  \right\vert
\end{align}%
\begin{align}
\mathbb{L}_{ab}  &  =\int\frac{d^{2}\alpha}{\pi}\left[  \int\frac{d^{2}w}{\pi
}\int\frac{d^{2}\alpha^{\prime}}{\pi}e^{-\left(  \frac{m}{\sqrt{2}\left\vert
\mathbf{a}\right\vert }\right)  ^{2}\left[  \left(  \alpha+\alpha^{\ast
}\right)  -B\right]  ^{2}+D}e^{\xi\varrho\left(  \alpha^{\prime}%
+\alpha^{\prime\ast}\right)  }\left\vert f\left(  \xi\right)  \right\vert
^{2}\left(
\begin{array}
[c]{c}%
\alpha^{\prime2}\\
\alpha^{\ast\prime2}%
\end{array}
\right)  _{\left(  1\right)  ab}\right.  \times\tag{35b}\\
&  \times\left.  e^{\frac{\left\vert w\right\vert ^{2}}{4}}e^{\frac{i}%
{2}\left[  \left(  \alpha-\alpha^{\prime}\right)  w^{\ast}+\left(
\alpha^{\ast}-\alpha^{\ast\prime}\right)  w\right]  }\right]  \left\vert
\Psi^{A}\left(  \alpha\right)  \right\rangle \left\langle \Psi^{A}\left(
\alpha\right)  \right\vert
\end{align}%
\begin{align}
\mathbb{L}_{ab}  &  =\int\frac{d^{2}\alpha}{\pi}\left[  \int\frac{d^{2}w}{\pi
}\int\frac{d^{2}\alpha^{\prime}}{\pi}e^{-\left(  \frac{m}{\sqrt{2}\left\vert
\mathbf{a}\right\vert }\right)  ^{2}\left[  \left(  \alpha+\alpha^{\ast
}\right)  -B\right]  ^{2}+D}e^{\xi\varrho\left(  \alpha^{\prime}%
+\alpha^{\prime\ast}\right)  }\left\vert f\left(  \xi\right)  \right\vert
^{2}\left(
\begin{array}
[c]{c}%
\alpha^{\prime2}\\
\alpha^{\ast\prime2}%
\end{array}
\right)  _{\left(  3/2\right)  ab}\right.  \times\tag{35c}\\
&  \times\left.  e^{\frac{\left\vert w\right\vert ^{2}}{4}}e^{\frac{i}%
{2}\left[  \left(  \alpha-\alpha^{\prime}\right)  w^{\ast}+\left(
\alpha^{\ast}-\alpha^{\ast\prime}\right)  w\right]  }\right]  \left\vert
\Psi_{3/4}\left(  \alpha\right)  \right\rangle \left\langle \Psi_{3/4}\left(
\alpha\right)  \right\vert
\end{align}%
\begin{align}
\mathbb{L}_{ab}  &  =\int\frac{d^{2}\alpha}{\pi}\left[  \int\frac{d^{2}w}{\pi
}\int\frac{d^{2}\alpha^{\prime}}{\pi}e^{-\left(  \frac{m}{\sqrt{2}\left\vert
\mathbf{a}\right\vert }\right)  ^{2}\left[  \left(  \alpha+\alpha^{\ast
}\right)  -B\right]  ^{2}+D}e^{\xi\varrho\left(  \alpha^{\prime}%
+\alpha^{\prime\ast}\right)  }\left\vert f\left(  \xi\right)  \right\vert
^{2}\left(
\begin{array}
[c]{c}%
\alpha^{\prime2}\\
\alpha^{\ast\prime2}%
\end{array}
\right)  _{\left(  1/2\right)  ab}\right.  \times\tag{35d}\\
&  \times\left.  e^{\frac{\left\vert w\right\vert ^{2}}{4}}e^{\frac{i}%
{2}\left[  \left(  \alpha-\alpha^{\prime}\right)  w^{\ast}+\left(
\alpha^{\ast}-\alpha^{\ast\prime}\right)  w\right]  }\right]  \left\vert
\Psi_{1/4}\left(  \alpha\right)  \right\rangle \left\langle \Psi_{1/4}\left(
\alpha\right)  \right\vert
\end{align}%
\begin{align}
L_{ab}  &  =\int\frac{d^{2}\alpha}{\pi}\left[  \int\frac{d^{2}w}{\pi}\int
\frac{d^{2}\alpha^{\prime}}{\pi}e^{-\left(  \frac{m}{\sqrt{2}\left\vert
\mathbf{a}\right\vert }\right)  ^{2}\left[  \left(  \alpha+\alpha^{\ast
}\right)  -B\right]  ^{2}+D}e^{\xi\varrho\left(  \alpha^{\prime}%
+\alpha^{\prime\ast}\right)  }\left\vert f\left(  \xi\right)  \right\vert
^{2}\left(
\begin{array}
[c]{c}%
\alpha^{\prime}\\
\alpha^{\ast\prime}%
\end{array}
\right)  _{\left(  2\right)  ab}\right.  \times\tag{36a}\\
&  \times\left.  e^{\frac{\left\vert w\right\vert ^{2}}{4}}e^{\frac{i}%
{2}\left[  \left(  \alpha-\alpha^{\prime}\right)  w^{\ast}+\left(
\alpha^{\ast}-\alpha^{\ast\prime}\right)  w\right]  }\right]  \left\vert
\Psi^{S}\left(  \alpha\right)  \right\rangle \left\langle \Psi^{S}\left(
\alpha\right)  \right\vert \nonumber
\end{align}%
\begin{align}
L_{ab}  &  =\int\frac{d^{2}\alpha}{\pi}\left[  \int\frac{d^{2}w}{\pi}\int
\frac{d^{2}\alpha^{\prime}}{\pi}e^{-\left(  \frac{m}{\sqrt{2}\left\vert
\mathbf{a}\right\vert }\right)  ^{2}\left[  \left(  \alpha+\alpha^{\ast
}\right)  -B\right]  ^{2}+D}e^{\xi\varrho\left(  \alpha^{\prime}%
+\alpha^{\prime\ast}\right)  }\left\vert f\left(  \xi\right)  \right\vert
^{2}\left(
\begin{array}
[c]{c}%
-\alpha^{\prime}\\
-\alpha^{\ast\prime}%
\end{array}
\right)  _{\left(  1\right)  ab}\right.  \times\tag{36b}\\
&  \times\left.  e^{\frac{\left\vert w\right\vert ^{2}}{4}}e^{\frac{i}%
{2}\left[  \left(  \alpha-\alpha^{\prime}\right)  w^{\ast}+\left(
\alpha^{\ast}-\alpha^{\ast\prime}\right)  w\right]  }\right]  \left\vert
\Psi^{A}\left(  \alpha\right)  \right\rangle \left\langle \Psi^{A}\left(
\alpha\right)  \right\vert \nonumber
\end{align}
and in the asymmetric case (34)%
\begin{align}
L_{ab}  &  =\int\frac{d^{2}\alpha}{\pi}\left[  \int\frac{d^{2}w}{\pi}\int
\frac{d^{2}\alpha^{\prime}}{\pi}e^{-\left(  \frac{m}{\sqrt{2}\left\vert
\mathbf{a}\right\vert }\right)  ^{2}\left[  \left(  \alpha+\alpha^{\ast
}\right)  -B\right]  ^{2}+D}e^{\xi\varrho\left(  \alpha^{\prime}%
+\alpha^{\prime\ast}\right)  }\left\vert f\left(  \xi\right)  \right\vert
^{2}\left(
\begin{array}
[c]{c}%
\alpha^{\prime}\\
\alpha^{\ast\prime}%
\end{array}
\right)  _{\left(  m+n\right)  ab}\right.  \times\tag{37}\\
&  \times\left.  e^{\frac{\left\vert w\right\vert ^{2}}{4}}e^{\frac{i}%
{2}\left[  \left(  \alpha-\alpha^{\prime}\right)  w^{\ast}+\left(
\alpha^{\ast}-\alpha^{\ast\prime}\right)  w\right]  }\right]  \left\vert
\Psi_{m}\left(  \alpha\right)  \right\rangle \left\langle \Psi_{n}\left(
\alpha^{\prime}\right)  \right\vert
\end{align}
with $m,n=1/4,3/4$ $\left(  m\neq n\right)  .$ The important point to remark
here is that there exist two different reconstructions for the operator
$L_{ab}$ in the full Hilbert space $\mathcal{H}$ with the same $\lambda$: a
diagonal representation and an asymmetric one. However, both representations
are not equivalent: for the asymmetric one the basic states involved are one
half than in the diagonal case and the eigenspinor $\left(
\begin{array}
[c]{c}%
\alpha^{\prime}\\
\alpha^{\ast\prime}%
\end{array}
\right)  _{\left(  1\right)  ab}$has plus sign (compare expressions (37) and
(36b)) . Then, as was pointed out for the authors of reference [35], in this
case the asymmetric representation is absolutely necessary to describe
completely the physical system (and also to reconstruct conveniently the operators).

Finally from the above expressions, the Gram-Schmidt operators can be easily
constructed $\left(  \lambda=1/2,1,3/2,2\right)  :$%
\[
\mathbb{G}=\int\frac{d^{2}\alpha}{\pi}\left[  \int\frac{d^{2}w}{\pi}%
e^{\frac{\left\vert w\right\vert ^{2}}{4}}e^{\frac{i}{2}\left[  \left(
\alpha-\alpha^{\prime}\right)  w^{\ast}+\left(  \alpha^{\ast}-\alpha
^{\ast\prime}\right)  w\right]  }\right]  \left\vert \Psi_{\lambda/2}\left(
\alpha\right)  \right\rangle \left\langle \Psi_{\lambda/2}\left(
\alpha\right)  \right\vert
\]%
\[
G=\int\frac{d^{2}\alpha}{\pi}\left[  \int\frac{d^{2}w}{\pi}e^{\frac{\left\vert
w\right\vert ^{2}}{4}}e^{\frac{i}{2}\left[  \left(  \alpha-\alpha^{\prime
}\right)  w^{\ast}+\left(  \alpha^{\ast}-\alpha^{\ast\prime}\right)  w\right]
}\right]  \left\vert \Psi_{\lambda/2}\left(  \alpha\right)  \right\rangle
\left\langle \Psi_{\lambda/2}\left(  \alpha\right)  \right\vert
\]
and in the asymmetric case (34)%
\[
G=\int\frac{d^{2}\alpha}{\pi}\left[  \int\frac{d^{2}w}{\pi}e^{\frac{\left\vert
w\right\vert ^{2}}{4}}e^{\frac{i}{2}\left[  \left(  \alpha-\alpha^{\prime
}\right)  w^{\ast}+\left(  \alpha^{\ast}-\alpha^{\ast\prime}\right)  w\right]
}\right]  \left\vert \Psi_{m}\left(  \alpha\right)  \right\rangle \left\langle
\Psi_{n}\left(  \alpha^{\prime}\right)  \right\vert
\]
where we solve naturally the identity in each (sub) Hilbert space, as is
required by, also in the non-diagonal case.

Summarizing explicitly the main results of this Section,

i) the Mp$\left(  2\right)  $ is the primary group that \ acts unitarily on
$\mathcal{H}$(and the same for the representation $\overline{\mathcal{H}}$)

i) In this model, the \textit{basic} CS are the fundamental solutions
generated by the most simplest non-degenerate supermetric where the
supercoordinates are the fields of the theory.

ii) These basic states have spin $\lambda$=1/4 and $\lambda$=3/4 and are not
physically observables.

ii) The basic CS generate a map (i.e. $g_{ab}$) that relates the operators
$L_{ab}$ and $\mathbb{L}_{ab}$ $\in Mp\left(  2\right)  $ to the specific
subspace of the full Hilbert space where these CS\ live.

iii) The \textit{physical states} are nothing more that tomographic
representations or quasiprobabilities in the sense that are the mean of the
operators $L_{ab}$ and $\mathbb{L}_{ab}$ $\in Mp\left(  2\right)  $(that
forms, with the dotted representation, the double covering of $SL\left(
2C\right)  $) with respect to the basic CS solution of the superwave equation
(given by expressions (32) );

ii) The representations for the operators $L_{ab}$ and $\mathbb{L}_{ab}$ $\in
Mp\left(  2\right)  $ (expressions (35-37)) are particular cases of a more
general kind of representations for operators recently proposed by Klauder and
Skagerstam in [35];

iii) the set of physical states are labeled by the total spin $\lambda$ and
the associated "eigenspinors"as described in expressions (33-34);

iv) in the best reconstruction formulas for $L_{ab}$ and $\mathbb{L}_{ab}$
(reliable in the sense given in [34]) the basic CS\ involved in such formulas
span all the Hilbert space $\mathcal{H=H}_{1/4}\oplus\mathcal{H}_{3/4}$;

v) there exist two types of \textit{non-equivalent} reconstruction formulas
for $L_{ab}$ and $\mathbb{L}_{ab}$: a diagonal representation and an
asymmetric one.

vi) from the previous point we conclude (in full agreement with the claim in
[35]) that the asymmetric representation is absolutely necessary in order to
describe completely the physical system (physic states).

\section{Discussion}

The proposal for the choice of model with a underlying basic structure starts
from the very early. Today, the large effective group given by the standard
model (multiplicity in the representations and the different coupling
constants) stimulates from time ago the search of such models. As we saw in
the first part of this work and other references, is that starting from the
most simplest non-degenerate supermetric where the supercoordinates are the
fields of the theory and retaining the original form of the fundamental
geometrical operators (namely Lagrangian or Hamiltonian) the physical states
obtained are constructed from the the basic ones by mean operators that
characterize the most fundamental symmetries of the spacetime.

The situation is more or less clear: although the supersymmetry is broken, the
physical states are localized in the "even" part of the manifold due the
metric coefficients of a non-degenerate supermetric. The physical states are
composed by most fundamental (non-observable) basic states. Operators
belonging to the metaplectic group (the most fundamental covering group of the
SL(2C)) lead, due a map produced by the basic CS, the observable spectrum of
physical states. This fact is clearly important as the "cornerstone" of a new
realistic composite model of particles based in coherent states where the
spacetime symmetry is directly connected with the physical spectrum.

\section{Concluding remarks}

In the present paper we have analyzed from the point of view of the symmetries
and the obtained vacuum solutions the superspace N=1 non-degenerate metric
proposed by Volkov and Pashnev in[9]. This particular model, although its high
simplicity, present a much richer structure than the others degenerate
standard superspaces because it contains the complex parameters $\mathbf{a}$
and $\mathbf{a}^{\ast}$ that make it different. The important role played by
the complex parameters $\boldsymbol{a}$ and $\boldsymbol{a}^{\ast}$ can be
resumed as follows:

i)the $\mathbb{C}$-parameters $\mathbf{a}$ and $\mathbf{a}^{\ast}$ fix the
field in a specific sector of the even part ($B_{L,0}$) of the supermanifold;

ii) these parameters, that are responsible of the non trivial part of the
model, break the chiral symmetry of the field solution. The chiral symmetry is
restored when the metric in question becomes degenerate in the limit
$\left\vert \mathbf{a}\right\vert \rightarrow\infty$ ( with all other
parameters of the model fixed);

iii) the fields remain attached in a specific region of the spacetime when the
supersymmetry of the model is completely broken, even if all the fermions are
switched off;

iv) we have analyzed and compared from the point of view of the obtained
solutions the superspace $\left(  1\mid2\right)  $ with the particular
superspace $\left(  1,3\mid1\right)  $ proposed by Volkov and Pashnev [9,11],
compactified to one dimension and restricted to the pure time-dependent case.
The possibility that the non-degenerate superspace $\left(  1,3\mid1\right)  $
with extended line element is reduced to the superspace $\left(
1\mid2\right)  $ is subject to the condition $\left\vert \mathbf{a}\right\vert
\rightarrow\infty$. The fermionic part of both superspaces is mapped one to
one by mean of a suitable definition of the fermionic variables and coefficients.

From the geometrical and group theoretical point of view the results are the following:

v) the supermetric can be derived from a gauge theory of supergravity based in
the $OSP(1/4)$ group;

vi) the complex parameters $\mathbf{a}$ and $\mathbf{a}^{\ast}$ play similar
role that the cosmological constant $\Lambda$ in the ordinary spacetime
models. Then, add a $\Lambda$ constant by hand is not necessary in this
supersymmetric model;

vii) a new generalization of the Volkov-Pashnev superparticle is presented for
N%
$>$%
1 supersymmetry where the first order fermionic term appear explicitly in the action.

In comparison with the 5-dimensional gravity plus cosmological constant of
refs.[14], the simple supersymmetric model under analysis here (now with
$n-$extra bosonic coordinates) has the following advantages:

viii) for $n=0$ the model, although is a very good candidate for a confinement
mechanism with natural breaking of the chiral symmetry in high energy physics
(e.g.:instanton liquid models, etc), cannot be compared directly with the
Randall-Sundrum model because the localization of the fields are not in the
bosonic extra-dimension (the physic are different in both cases).

For $n=1$

ix) the mechanism of localization of the fields in the bosonic 4-dimensional
part of the supermanifold does not depends on the cosmological constant;

x) the fields attached are Gaussian type solutions (7) very well defined
physical state in a Hilbert space from the mathematical point of view,
contrarily to the case $u\left(  y\right)  =ce^{-H\left\vert y\right\vert }$
given in [16];

xi) not additional and/or topological structures that break the symmetries of
the model (i.e. reflection $Z_{2}$-symmetry) are required to attach the
fields: the natural structure of the superspace produces this effect through
the $\mathbb{C}$-parameters $\mathbf{a}$ and $\mathbf{a}^{\ast}$.

From the point of view of the obtained spectrum of physic states we explicitly
shown that

i) the Mp$\left(  2\right)  $ is the primary group that \ acts unitarily on
$\mathcal{H}$(and the same for the representation $\overline{\mathcal{H}}$)

i) In this model, the \textit{basic} CS are the fundamental solutions
generated by the most simplest non-degenerate supermetric where the
supercoordinates are the fields of the theory.

ii) These basic states have spin $\lambda$=1/4 and $\lambda$=3/4 and are not
physically observables.

ii) The basic CS generate a map (i.e. $g_{ab}$) that relates the operators
$L_{ab}$ and $\mathbb{L}_{ab}$ $\in Mp\left(  2\right)  $ to the specific
subspace of the full Hilbert space where these CS\ live.

iii) The \textit{physical states} are nothing more that tomographic
representations or quasiprobabilities in the sense that are the mean of the
operators $L_{ab}$ and $\mathbb{L}_{ab}$ $\in Mp\left(  2\right)  $(that
forms, with the dotted representation, the double covering of $SL\left(
2C\right)  $) with respect to the basic CS solution of the superwave equation
(given by expressions (32) );

ii) The representations for the operators $L_{ab}$ and $\mathbb{L}_{ab}$ $\in
Mp\left(  2\right)  $ (expressions (35-37)) are particular cases of a more
general kind of representations for operators recently proposed by Klauder and
Skagerstam in [35];

iii) the set of physical states are labeled by the total spin $\lambda$ and
the associated "eigenspinors"as described in expressions (33-34);

iv) in the best reconstruction formulas for $L_{ab}$ and $\mathbb{L}_{ab}$
(reliable in the sense given in [34]) the basic CS\ involved in such formulas
span all the Hilbert space $\mathcal{H=H}_{1/4}\oplus\mathcal{H}_{3/4}$;

v) there exist two types of \textit{non-equivalent} reconstruction formulas
for $L_{ab}$ and $\mathbb{L}_{ab}$: a diagonal representation and an
asymmetric one.

vi) the asymmetric representation is absolutely necessary in order to describe
completely the physical system (physic states).

\section{Acknowledgements}

I am very thankful to the professor E. A. Ivanov for his interest and give me
several remarks and references on the descriptions of the superspaces. I
appreciate deeply Professor Alexander Dorokhov and Sergey Molodzov's for
useful discussions concerning to the topological properties of the vacuum and
the confinement mechanism I am very grateful to the Professor Jose Luis Boya
for useful discussions and all the the people of the Department of Theoretical
Physics of the University of Zaragoza for they hospitality and support, were
this work was finished.

\section{Appendix}

The dynamics of the $\left\vert \Psi\right\rangle $ fields , in the
representation that we are interested in, can be simplified considering these
fields as coherent states in the sense that are eigenstates of $a^{2}$ [19]%

\begin{align}
\left\vert \Psi_{1/4}\left(  0,\xi,q\right)  \right\rangle  &  =\overset
{+\infty}{\underset{k=0}{\sum}}f_{2k}\left(  0,\xi\right)  \left\vert
2k\right\rangle =\overset{+\infty}{\underset{k=0}{\sum}}f_{2k}\left(
0,\xi\right)  \frac{\left(  a^{\dag}\right)  ^{2k}}{\sqrt{\left(  2k\right)
!}}\left\vert 0\right\rangle \tag{A1}\\
\left\vert \Psi_{3/4}\left(  0,\xi,q\right)  \right\rangle  &  =\overset
{+\infty}{\underset{k=0}{\sum}}f_{2k+1}\left(  0,\xi\right)  \left\vert
2k+1\right\rangle =\overset{+\infty}{\underset{k=0}{\sum}}f_{2k+1}\left(
0,\xi\right)  \frac{\left(  a^{\dagger}\right)  ^{2k+1}}{\sqrt{\left(
2k+1\right)  !}}\left\vert 0\right\rangle \nonumber
\end{align}
From a technical point of view these states are a one mode squeezed states
constructed by the action of the generators of the SU(1,1) group over the
vacuum. For simplicity, we will take all normalization and fermionic
dependence or possible CS fermionic realization, into the functions $f\left(
\xi\right)  $. Explicitly at t=0
\begin{equation}%
\begin{array}
[c]{c}%
\left\vert \Psi_{1/4}\left(  0,\xi,q\right)  \right\rangle =f\left(
\xi\right)  \left\vert \alpha_{+}\right\rangle \\
\left\vert \Psi_{3/4}\left(  0,\xi,q\right)  \right\rangle =f\left(
\xi\right)  \left\vert \alpha_{-}\right\rangle
\end{array}
\tag{A2}%
\end{equation}
where $\left\vert \alpha_{\pm}\right\rangle $ are the CS basic states in the
subspaces $\lambda=\frac{1}{4}$ and $\lambda$ =$\frac{3}{4}$ of the full
Hilbert space. In the case of the physical state that we are interested in, we
used the HW realization for the states $\Psi$%
\begin{equation}
\left\vert \Psi\right\rangle =\frac{f\left(  \xi\right)  }{2}\left(
\left\vert \alpha_{+}\right\rangle +\left\vert \alpha_{-}\right\rangle
\right)  =f\left(  \xi\right)  \left\vert \alpha\right\rangle \tag{A3}%
\end{equation}
where, however, the linear combination of the states $\left\vert \alpha
_{+}\right\rangle $ and $\left\vert \alpha_{-}\right\rangle $ span now the
full Hilbert space (dense) being the correspond $\lambda$ to the CS basis .The
"square" state at t=0 are%
\begin{align}
g_{ab}\left(  0\right)   &  =\left\langle \Psi\left(  0\right)  \right\vert
L_{ab}\left\vert \Psi\left(  0\right)  \right\rangle =\left\langle \Psi\left(
0\right)  \right\vert \left(
\begin{array}
[c]{c}%
a\\
a^{+}%
\end{array}
\right)  _{ab}\left\vert \Psi\left(  0\right)  \right\rangle \tag{A4}\\
&  =f^{\ast}\left(  \xi\right)  f\left(  \xi\right)  \left(
\begin{array}
[c]{c}%
\alpha\\
\alpha^{\ast}%
\end{array}
\right)  _{ab}\nonumber
\end{align}

The algebra (topological information of the group manifold) is "mapped" over
the spinors solutions through the eigenvalues $\alpha$ and $\alpha^{\ast}$.
Notice that the constants $c_{1}^{\prime}$ $c_{2}^{\prime}$ in the exponential
functions \ in expressions (10) and (11) can be easily determined as functions
of the frequency $\omega$ as in ref.[19] for the Schr\"{o}dinger equation.

\section{References}

[1] V. P. Akulov, V. A. Soroka and D. V. Volkov, JETP Lett. \textbf{22}, 396
(1975), in Russian; Theor. Math. Phys.\textbf{ 31}, 12 (1977); B. Zumino, CERN
Preprint TH2120 (1975); J. Wess and B. Zumino, Phys. Lett. B \textbf{66}, 361 (1977).

[2] R. Arnowitt, P. Nath and B. Zumino, Nucl. Phys. B \textbf{56}, 81 (1975).

[3] V. Ogievetsky and E. Sokatchev, Phys. Lett. B \textbf{79}, 222 (1978).

[4] W. Siegel and S. J. Gates Jr., Nucl. Phys. B \textbf{147}, 77 (1978).

[5] I. L. Buchbinder and S. M. Kuzenko: \textit{Ideas and Methods of
Supersymmetry and Supergravity} (Philadelphia, PA:IOP Publishing, 1995); B. De
Witt: \textit{Supermanifolds} (Cambridge University Press, Cambridge, 1984).

[6] S.B. Giddings and A. Maharana, Phys. Rev. D \textbf{73}, 126003 (2006); K.
Koyama, arXiv: astro-ph/0706.1557.

[7] T. Curtright et al., JHEP 0704.020 (2007), and references therein.

[8] S. Bellucci et al., Phys. Rev. D \textbf{66}, 086001 (2002), and
references therein.

[9] A.I. Pashnev and D. V. Volkov, Teor. Mat. Fiz., Tom. 44, No.\textbf{ 3},
321 (1980) [in Russian], Andrzey Frydryzak, \textit{Lagrangian models of
particles with spin}, arXiv:hep-th/9601020v1 (6 Jan 1996).

[10] D. J. Cirilo-Lombardo, Foundations of Physics, Vol. 37, No. \textbf{6},
(2007), and references therein.

[11] D. J. Cirilo-Lombardo, Rom. Journal of Physics \textbf{7-8}, Vol 50, 875,
(2005); Elem. Particles and Nucl. Lett.\textbf{6, }Vol. 3, 416 (2006);
Hadronic J.\textit{ }\textbf{29,} 355 (2006); Elem. Particles and Nucl.
Lett.\textbf{3, }Vol. 4, 239 (2007)..

[12] V. P. Akulov and D. V. Volkov, Theor. Math. Phys. 41, 939 (1979).

[13] V. P. Akulov and D. V. Volkov, Theor. Math. Phys. 42, 10 (1980).

[14] B. Bajc and G. Gabadadze, Phys. Lett. B \textbf{474}, 282 (2000), and
references therein.

[15] O. Bertolami et al., Int. J. Mod. Phys. \textbf{A6}, 4149 (1991).

[16] D. J. Cirilo-Lombardo, in preparation.

[17] D. Alfaro, S. Fubini and G. Furlan, Nuovo Cimento A \textbf{34}, 569 (1976).

[18] R. Arnowitt and P. Nath , Phys. Rev. D \textbf{15}, 1033 (1977).

[19] J. R. Klauder and B. S. Skagerstam: \textit{Coherent States} (World
Scientific, Singapore, 1985).

[20] N. Arkani-Hamed et al., Phys. Lett. B \textbf{429}, 263 (1998); Phys.
Rev. D \textbf{59}, 0860 (1999).

[21] V. A. Rubakov and M. E. Shaposhnikov, Phys. Lett. B \textbf{125}, 136 (1983).

[22] G. Dvali and M. Shifman, Phys. Lett. B \textbf{396}, 64 (1997); Nucl.
Phys. B \textbf{504}, 127 (1996).

[23] J. Polchinski, Phys. Rev. Lett. \textbf{75, }4724, (1995)\textbf{.}

[24] M. Gogberashvili, hep-ph/9812296; hep-ph/9812365.

[25] L. Randall and R. Sundrum, Phys. Rev. Lett. \textbf{83, }3370 (1999);
Phys. Rev. Lett. \textbf{83, }4690 (1999).

[26] M. Gabella et al., Phys. Rev. D \textbf{76}, 055001 (2007), and
references therein.

[27] F. Constantinescu, J. Phys. A: Math. Gen. \textbf{38}, 1385 (2005).

[28] S. S. Sannikov: Non-compact symmetry group of a quantum oscillator.
\textit{Zh.E.T.F.} \textbf{49}, 1913 (1965), [in Russian].

[29] T. Goldman et al., Phys. Lett. B \textbf{171}, 217 (1986); M.M. Nieto,
private communication.

[30] Hans-J\"{u}rgen Treder and Wolfgang Yourgrou, Fundations of Physics,
\textbf{8}, 695 (1978), and references therein.

[31] H. C. Corben, Fundations of Physics, \textbf{10}, 181 (1980), and
references therein.

[32] P. van Nieuwenhuizen, Physics Reports \textbf{68}, 189 (1981), and
references therein.

[33] Jose A. Azcarraga and Jerzy Lukierski, Z. Phys. C., \textbf{30}, 221
(1986), and references therein.

[34] G. Marmo et al., Jornal of Physics: Conference Series \textbf{87}, 012013 (2007).

[35] John R. Klauder and Bo-Sture K. Skagerstam, J.Phys.A: Math. Theor.
\textbf{40,} 2013 (2007).

[36] R. Simon and N. Mukunda, The Two-dimensional Symplectic and Metaplectic
Groups and Their Universal Cover in \textit{Symmetries in science VI: from the
rotation groups to the quantum algebras}, edited by B. Gruber (Plenum, New
York, 1994).

[37] John R. Klauder and E. C. G. Sudarshan, \textit{Fundamentals of Quantum
Optics} (New York: Benjamin, 1968).

[38] see fo example D. P. Sorokin and D. V. Volkov: (Anti) commuting spinors
and supersymmetric dynamics of semions. \textit{Nucl. Phys.} \textbf{B409},
547 (1993); D. P. Sorokin: The Heisenberg algebra and spin. \textit{Fortschr.
Phys.} \textbf{50}, 724 (2002).

\end{document}